\documentclass{amsart}
\usepackage{amsmath,amssymb,amsfonts,epsfig}

\newtheorem{theorem}{Theorem}[section]
\newtheorem{proposition}[theorem]{Proposition}

\newtheorem{corollary}[theorem]{Corollary}

\theoremstyle{definition}

\theoremstyle{remark}

\numberwithin{equation}{section}

\def\DJ{{\hbox{D\kern-.8em\raise.15ex\hbox{--}\kern.35em}}}
\def\DJo{\DJ okovi\'c}

\font\germ=eufm10

\def\su{{\mbox{\germ su}}}

\def\be{{\beta}}

\def\vf{{\varphi}}
\def\sig{{\sigma}}
\def\bC{{\bf C}}

\def\bZ{{\bf Z}}
\def\pA{{\mathcal A}}
\def\pD{{\mathcal D}}
\def\pH{{\mathcal H}}
\def\pM{{\mathcal M}}
\def\pP{{\mathcal P}}
\def\pS{{\mathcal S}}

\def\diag{{\rm diag}}

\def\tr{{\rm \;tr}}

\def\SU{{\mbox{\rm SU}}}
\def\Un{{\mbox{\rm U}}}

\let\la=\langle  \let\ra=\rangle

\begin{document}

\title[ Multigraded Poincar\'{e} series for two qubits ]
{Multigraded Poincar\'{e} series for mixed states of two qubits
and the boundary of the set of separable states} 

\author[D.\v{Z}. \DJo{}]
{Dragomir \v{Z}. \DJo{}}

\address{Department of Pure Mathematics, University of Waterloo,
Waterloo, Ontario, N2L 3G1, Canada}

\email{djokovic@uwaterloo.ca}


\thanks{The author was supported in part by the NSERC
Grant A-5285.}

\date{}

\begin{abstract}
Let $\pM$ be the set of mixed states and $\pS$ the set of
separable states of the two-qubit system. Its
Hilbert space is the tensor product $\pH=\bC^2\otimes\bC^2$,
and the group of local unitary transformations is
$G=\SU(2)\times\SU(2)$ (ignoring the overall phase factor).
Let $\pP$ be the algebra of real polynomial functions on the
space of all hermitian operators of trace 1 on $\pH$.
Let $\pP^G\subseteq\pP$ be the subalgebra of $G$-invariants.
We compute its multigraded Poincar\'{e} series and verify that it
is consistent with Makhlin's list of 18 invariants.
By using the recent result of Augusiak et al. we describe the
boundary of $\pS$ and show that its intersection
with the (relative) interior of $\pM$ is a smooth manifold.
\end{abstract}

\maketitle

\section{Introduction}

Let $\bC^n$ be the Hilbert space (of column vectors)
with the inner product $\la x_1,x_2 \ra=x_1^\dag x_2$ and
let $M_n$ be the algebra of complex $n\times n$ matrices.
If $X,Y\in M_n$ then their inner product is
defined by $\la X_1,X_2 \ra=\tr(X_1^\dag X_2)$.
Let $\su(n)$ denote the Lie algebra of $\SU(n)$, it consists
of all traceless skew-hermitian matrices in $M_n$.

The quantum system that we consider is bipartite.
Both parties are qubits, $\bC^2$.
It will be convenient to identify $\bC^4$ with the tensor product
$\pH=\bC^2\otimes_\bC \bC^2$, and $M_4$ with $M_2\otimes_\bC M_2$.
For $X_1,X_2,Y_1,Y_2\in M_2$, the inner product of
$X_1\otimes Y_1$ and $X_2\otimes Y_2$ is given by
\[ \la X_1\otimes Y_1,X_2\otimes Y_2 \ra=
\la X_1,X_2 \ra\cdot \la Y_1,Y_2 \ra. \]
It follows that $(X\otimes Y)^\dag=X^\dag\otimes Y^\dag$ and,
in particular, the tensor product of hermitian matrices is
again a hermitian matrix.

The space of hermitian matrices $H_4\subseteq M_4$ is the direct sum of
the $1D$ real space spanned by the identity matrix, $I_4$,
and the space $H_{4,0}=i\su(4)$. The space
$H_{4,0}$ is the direct sum of three real subspaces:
$V_1=H_{2,0}\otimes I_2$, $V_2=I_2\otimes H_{2,0}$,
and $V_3=H_{2,0}\otimes H_{2,0}$.
Any mixed state, $\rho$, of our quantum system can be written
uniquely as the sum of four components;
\begin{equation} \label{stanje}
\rho=\frac{1}{4}I_4+X\otimes I_2+I_2\otimes Y+Z,
\end{equation}
where $X,Y\in H_{2,0}$ and $Z\in V_3$.

Denote by $G$ the group of local unitary transformations,
$\SU(2)\times\SU(2)$, where we ignore the overall phase factor.
Note that $G$ acts on $M_4$ in the usual manner:
$(g,Z)\to gZg^\dag$ and stabilizes the real subspaces $H_4$
and $H_{4,0}$. Moreover, each of the subspaces
$V_1$, $V_2$, $V_3$ is a simple $G$-module.

Let $\pP$ denote the algebra of real valued polynomial
functions on $H_{4,0}$ and
$\pP^G$ the subalgebra of $G$-invariant functions.
Note that $\pP^G$ inherits the $\bZ$-gradation from $\pP$.
The homogeneous polynomials of $\pP^G$ may be used to construct
measures of entanglement of our quantum system.

The knowledge of $\pP^G$ is important because of
the following well known fact: Two states, say $\rho_1$ and
$\rho_2$, belong to different $G$-orbits iff
there exists $f\in\pP^G$ such that $f(\rho_1)\ne f(\rho_2)$.
If this holds true for a subalgebra $\pA\subseteq\pP^G$,
then we say that $\pA$ is \emph{complete}.

The Poincar\'{e} series (also known as the Hilbert series)
of $\pP^G$ was computed by M. Grassl et al. \cite[Section VI]{GRB}.
We have
\[ P(z)=\sum_{d=0}^\infty \dim(\pP_d^G) z^d, \]
where $\pP_d^G$ is the space of homogeneous polynomial
$G$-invariants of degree $d$. This Poincar\'{e} series is a rational
function of the variable $z$:
\begin{equation} \label{prosta}
P(z)=
{\frac {1-{z}^{2}-{z}^{3}+2\,{z}^{4}+2\,{z}^{5}+2\,{z}^{6}-{z}^{7}-{z}
^{8}+{z}^{10}}{ \left( 1-z \right) ^{9} \left( 1+z \right) ^{6}
 \left( 1+{z}^{2} \right) ^{2} \left( 1+z+{z}^{2} \right) ^{3}}}.
\end{equation}
We remark that in \cite{GRB} the formula contains an extra factor
$1-z$ in the denominator because the authors work with the space $H_4$
instead of $H_{4,0}$.

The Taylor expansion begins with
\begin{eqnarray*}
P(z) &=& 1+3\,{z}^{2}+2\,{z}^{3}+10\,{z}^{4}+7\,{z}^{5}+29\,{z}^{6}
+25\,{z}^{7}+73\,{z}^{8}+74\,{z}^{9}+172\,{z}^{10} \\
&& +187\,{z}^{11}+381\,{z}^{12}+431\,{z}^{13}+785\,{z}^{14}+920\,{z}^{15}
+1539\,{z}^{16}+1827\,{z}^{17} \\
&& +2878\,{z}^{18}+3441\,{z}^{19}+5151\,{z}^{20}+6185\,{z}^{21}
+8887\,{z}^{22}+10666\,{z}^{23}+\cdots
\end{eqnarray*}

Subsequently, Y. Makhlin \cite{YM} gave a simple construction of
18 invariants which generate a complete subalgebra of $\pP^G$.
This subalgebra is proper and he gives two additional invariants.
However, it is still not established whether this enlarged subalgebra
is in fact the whole algebra $\pP^G$.
(According to M. Grassl \cite{MG}
this can be proved by using the relations for
the invariant tensors of $\SU(2)$.)

As a complement to the two papers just mentioned,
we compute in Section \ref{mult} the $\bZ^3$-graded Poincar\'{e}
series of $\pP^G$. This multigraded series carries more
information about the invariants than the simply graded one.

Let us recall that a mixed state is called \emph{separable} if
it can be represented as a convex combination of product states,
and otherwise it is called \emph{entangled}.
Denote by $\pM$ the set of all states and by $\pS$
the set of all separable states.
It is well known that these two sets are compact and convex,
and have nonempty interior.
By $\partial\pM$ resp. $\partial\pS$ we denote the boundary of
$\pM$ and $\pS$, respectively.

In Section \ref{bound} we show that there is a natural way of
decomposing the boundary $\partial\pS$ into two pieces.
Each of these pieces is a portion of a real algebraic hypersurface
in the ambient affine space. In particular,
it follows from our results that the intersection of $\partial\pS$
with the relative interior of $\pM$ is a smooth manifold.

\section{ The multigraded Poincar\'{e} series }
\label{mult}

The direct decomposition $H_{4,0}=V_1\oplus V_2\oplus V_3$
induces a $\bZ^3$-gradation on $\pP$ which is preserved by the action
of $G$. If the space of homogeneous invariants of degrees
$d_1$, $d_2$, $d_3$ (with respect to the coordinates of the three
subspaces) has dimension $c$, then this fact is recorded in the
Taylor expansion of the multigraded Poincar\'{e} series by the
term $ct_1^{d_1}t_2^{d_2}t_3^{d_3}$.

\begin{theorem} \label{2-kubita}
The multigraded Poincar\'{e} series $P(t_1,t_2,t_3)$ of the
algebra $\pP^G$ of local unitary polynomial invariants of mixed
states of two qubits is the rational function, whose numerator
$N(t_1,t_2,t_3)$ and denominator $D(t_1,t_2,t_3)$ are given by:
\begin{eqnarray*}
N &=& 1-t_{{1}}{t_{{3}}}^{2}-t_{{2}}{t_{{3}}}^{2}+t_{{1}}t_{{2}}{t_{{3}}}^{2}
+t_{{1}}t_{{2}}{t_{{3}}}^{3}+{t_{{1}}}^{2}t_{{2}}{t_{{3}}}^{3}
+t_{{1}}{t_{{2}}}^{2}{t_{{3}}}^{3}+{t_{{1}}}^{2}{t_{{3}}}^{4}+
t_{{1}}t_{{2}}{t_{{3}}}^{4} \\
&& +{t_{{2}}}^{2}{t_{{3}}}^{4}-{t_{{1}}}^{3}t_{{2}}{t_{{3}}}^{5} 
-{t_{{1}}}^{2}{t_{{2}}}^{2}{t_{{3}}}^{5}-t_{{1}}{t_{{2}}}^{3}{t_{{3}}}^{5}  
-{t_{{1}}}^{2}t_{{2}}{t_{{3}}}^{6}-t_{{1}}{t_{{2}}}^{2}{t_{{3}}}^{6} \\
&& -{t_{{1}}}^{2}{t_{{2}}}^{2}{t_{{3}}}^{6}
-{t_{{1}}}^{2}{t_{{2}}}^{2}{t_{{3}}}^{7}
+{t_{{1}}}^{3}{t_{{2}}}^{2}{t_{{3}}}^{7}
+{t_{{1}}}^{2}{t_{{2}}}^{3}{t_{{3}}}^{7}
-{t_{{1}}}^{3}{t_{{2}}}^{3}{t_{{3}}}^{9}, \\
D &=&
 \left( 1-{t_{{1}}}^{2} \right)  \left( 1-{t_{{2}}}^{2} \right) 
 \left( 1-{t_{{3}}}^{2} \right)  \left( 1-t_{{1}}t_{{2}}t_{{3}}
 \right)  \left( 1-t_{{1}}{t_{{3}}}^{2} \right)
 \left( 1-t_{{2}}{t_{{3}}}^{2} \right) \left( 1-{t_{{3}}}^{3} \right) \\
&&  \left( 1-{t_{{1}}}^{2}{t_{{3}}}^{2} \right)
 \left( 1-{t_{{2}}}^{2}{t_{{3}}}^{2} \right) \left( 1-{t_{{3}}}^{4} \right).
\end{eqnarray*}
\end{theorem}

The theorem is proved by using the well-known Molien--Weyl formula.
Following the recipe from \cite{DK}, we obtain that
\[ P(t_1,t_2,t_3)=\frac{1}{(2\pi i)^3}
\int_{|z|=1} \int_{|y|=1} \int_{|x|=1} \vf(x,y,z,t_1,t_2,t_3)
\frac{ {\rm d}x }{x} \frac{ {\rm d}y }{y} \frac{ {\rm d}z }{z}, \]
where
\[ \vf(x,y,z,t_1,t_2,t_3)=\frac{(1-x^{-1})(1-y^{-1})(1-z^{-1})(1-y^{-1}z^{-1})}
{\psi(x,y,z,t_1,t_2,t_3)} \]
and
\begin{eqnarray*}
\psi &=& (1-t_1)(1-t_2)^2(1-t_3)^2
(1-t_1x)(1-t_1x^{-1})(1-t_3x)^2(1-t_3x^{-1})^2 \\
&& (1-t_3xy)(1-t_3xz)(1-t_3xyz)
(1-t_3xy^{-1})(1-t_3xz^{-1})(1-t_3xy^{-1}z^{-1})  \\
&& (1-t_3x^{-1}y)(1-t_3x^{-1}z)(1-t_3x^{-1}yz) \\
&& (1-t_3x^{-1}y^{-1})(1-t_3x^{-1}z^{-1})(1-t_3x^{-1}y^{-1}z^{-1})  \\
&& (1-t_2y)(1-t_2yz)(1-t_2y^{-1})(1-t_2y^{-1}z^{-1}) \\
&& (1-t_3y)(1-t_3yz)(1-t_3y^{-1})(1-t_3y^{-1}z^{-1}) \\
&& (1-t_2z)(1-t_2z^{-1})(1-t_3z)(1-t_3z^{-1}).
\end{eqnarray*}
The three integrations are to be performed over the unit circle in the
counterclockwise direction assuming that the $|t_i|<1$.
The computation was carried out by using Maple \cite{Map}.

After setting $t_1=t_2=t_3=z$ in $P(t_1,t_2,t_3)$, the numerator and
denominator acquire the common factor $(1+z^2)(1-z^3)$.
After cancellation, we obtain exactly the Eq. (\ref{prosta}),
i.e., we have $P(z,z,z)=P(z)$. 

By expanding $P(t_1,t_2,t_3)$ in the Taylor series, we find that

\begin{eqnarray*}
P &=&
1+{t_{{1}}}^{2}+{t_{{2}}}^{2}+{t_{{3}}}^{2}+{t_{{3}}}^{3}+
t_{{1}}t_{{2}}t_{{3}}+2\,{t_{{3}}}^{4}+2\,{t_{{2}}}^{2}{t_{{3}}}^{2}
+{t_{{2}}}^{4}+t_{{1}}t_{{2}}{t_{{3}}}^{2} \\
&& +2\,{t_{{1}}}^{2}{t_{{3}}}^{2}+{t_{{1}}}^{2}{t_{{2}}}^{2}
+{t_{{1}}}^{4}+{t_{{3}}}^{5}+{t_{{2}}}^{2}{t_{{3}}}^{3}
+2\,t_{{1}}t_{{2}}{t_{{3}}}^{3}
+t_{{1}}{t_{{2}}}^{3}t_{{3}}+{t_{{1}}}^{2}{t_{{3}}}^{3} \\
&& +{t_{{1}}}^{3}t_{{2}}t_{{3}}+3\,{t_{{3}}}^{6}+4\,{t_{{2}}}^{2}{t_{{3}}}^{4}
+2\,{t_{{2}}}^{4}{t_{{3}}}^{2}+{t_{{2}}}^{6}+2\,t_{{1}}t_{{2}}{t_{{3}}}^{4}
+t_{{1}}{t_{{2}}}^{2}{t_{{3}}}^{3}+t_{{1}}{t_{{2}}}^{3}{t_{{3}}}^{2} \\
&& +4\,{t_{{1}}}^{2}{t_{{3}}}^{4}+{t_{{1}}}^{2}t_{{2}}{t_{{3}}}^{3}
+4\,{t_{{1}}}^{2}{t_{{2}}}^{2}{t_{{3}}}^{2}+{t_{{1}}}^{2}{t_{{2}}}^{4}
+{t_{{1}}}^{3}t_{{2}}{t_{{3}}}^{2}+2\,{t_{{1}}}^{4}{t_{{3}}}^{2} \\
&& +{t_{{1}}}^{4}{t_{{2}}}^{2}+{t_{{1}}}^{6} +\cdots
\end{eqnarray*}

To make the connection with the notation in Makhlin's paper, we may
assume that the components of his vectors ${\bf s}$ and ${\bf p}$
are the linear coordinates on $V_1$ and $V_2$, respectively, and
the nine entries $\be_{ij}$ of his matrix $\hat{\be}$ are the
coordinates on $V_3$. Then it is easy
to check that his list of invariants agrees with the information
provided by the coefficients in the above Taylor expansion.
For instance, the cubic terms in the above expansion are
$t_3^3$ and $t_1t_2t_3$. They correspond to Makhlin's
invariants $I_1$ and $I_{12}$, respectively.

\section{ The boundary of the set of separable states }
\label{bound}

Using Makhlin's notation, we write the components $X,Y,Z$ in
Eq. (\ref{stanje}) as
\[ X=\frac{1}{2}\sum s_i\sig_i,\quad
Y=\frac{1}{2}\sum p_i\sig_i,\quad
Z=\sum_{i,j}\be_{ij}\sig_i\otimes\sig_j, \]
where the $\sig_i$'s are the Pauli matrices
\begin{equation*}
\sig_1=\left( \begin{array}{rr} 0&1\\1&0 \end{array} \right),\quad
\sig_2=\left( \begin{array}{rr} 0&-i\\i&0 \end{array} \right),\quad
\sig_3=\left( \begin{array}{rr} 1&0\\0&-1 \end{array} \right).
\end{equation*}
In \cite[Table 1]{YM} Makhlin lists his 18 invariants $I_1,\ldots,I_{18}$.
We shall need only 9 of them:
\begin{eqnarray*}
&& I_1=\det\hat{\be},\quad I_2=\tr(\hat{\be}^T\hat{\be}),
\quad I_3=\tr(\hat{\be}^T\hat{\be})^2, \\
&& I_4=\mathbf{s}^2,\quad I_5=[\mathbf{s}\hat{\be}]^2,\quad
I_7=\mathbf{p}^2,\quad I_8=[\hat{\be}\mathbf{p}]^2, \\
&& I_{12}=\mathbf{s}\hat{\be}\mathbf{p},\quad
I_{14}=e_{ijk}e_{lmn}s_ip_l\be_{jm}\be_{kn},
\end{eqnarray*}
where $e_{ijk}$ is the Levi--Civita symbol.

In a recent paper \cite{AHD} Augusiak et al. have shown that
$\pS$ is exactly the set of all states $\rho$ satisfying the
inequality $\det\rho^\Gamma\ge0$,
where $\Gamma$ is the operator of partial transposition,
say with respect to the second party.
Denote by $\pD$ the real algebraic hypersurface
in the affine space $(1/4)I_4+H_{4,0}$ given by the equation
$\det\rho=0$. Its image under $\Gamma$, which we denote by
$\pD^\Gamma$, is defined by the equation $\det \rho^\Gamma=0$.

Let us recall that $\partial\pM$ consists of all $\rho\ge0$
(with $\tr \rho=1$) having at least one zero eigenvalue, and so
$\partial\pM\subseteq\pD$.
This inclusion is in fact proper, i.e., $\partial\pM$ is only
a small portion of the entire hypersurface $\pD$. By the
Peres--Horodecki criterion \cite{HHH} we know that
$\pS=\pM\cap\pM^\Gamma$. Hence, $\partial\pS$ is the union
of two pieces:
\[ \partial\pS=\left( \pD \cap \pM^\Gamma \right) \cup
\left( \pD^\Gamma \cap \pM \right). \]
The first of these pieces has been mentioned in the paper
\cite{VDM} of Verstraete et al.

The polynomial function $\det\rho^\Gamma$ is an invariant of $G$.
We have found the following expression for it in terms of the
above Makhlin's invariants:
\begin{eqnarray*}
\det\rho^\Gamma &=& \frac{1}{256}-\frac{1}{32}(4I_2+I_4+I_7)
+\frac{1}{2}(4I_1+I_{12})+\frac{1}{16} \left( 32I_3-16I_5 \right. \\
&& \left. -16I_8-16I_{14}-16I_2^2+I_4^2+I_7^2
+8I_2I_4+8I_2I_7-2I_4I_7 \right) .
\end{eqnarray*}

There are many papers devoted to the study of the geometry
of the sets $\pM$ and $\pS$, e.g. \cite{GB,KZ,OR,SD,VDM,ZS}.
It may be of interest to study the boundary $\partial\pS$
and the determinantal hypersurface $\pD$ in more detail.
Let us say that a point $\rho_0\in\pD$ is
\emph{smooth} if the gradient $\nabla\det\rho$ does
not vanish at $\rho_0$, and otherwise it is a \emph{singular}
point of $\pD$.

The next proposition is valid for any $n$-dimensional complex
Hilbert space $\pH$. In this more general setting both $\pD$ and
$\pD^\Gamma$ are
hypersurfaces in the real affine space $(1/n)I_n+H_{n,0}$.

\begin{proposition} \label{sing}
A point $\rho_0\in\pD$ is a singular point of the hypersurface
$\pD$ if and only if $\rho_0$ has at least two zero eigenvalues.
\end{proposition}

\begin{proof}
Let us arrange the eigenvalues of $\rho_0$ in increasing order
\[ \lambda_1\le\lambda_2\le\cdots
\le\lambda_n=1-\sum_{k=1}^{n-1}\lambda_k \]
and observe that we must have $\lambda_n>0$.
Since the unitary group $\Un(n)$ preserves $\pD$ and maps
singular points to singular points, we may assume that $\rho_0$
is in fact the diagonal matrix
\[ \rho_0=\diag(\lambda_1,\lambda_2,\ldots,\lambda_{n-1},
1-\lambda_1-\cdots-\lambda_{n-1}). \]
For an arbitrary $\rho$, the determinant expansion has the form
\[ \det \rho=\rho_{11}\rho_{22}\cdots\rho_{n-1,n-1}
(1-\rho_{11}-\cdots-\rho_{n-1,n-1})+P, \]
where each term of the polynomial $P$ is at least quadratic in
the off diagonal entries of $\rho$. Hence in evaluating the
gradient of $\det\rho$ at the point $\rho_0$ the polynomial $P$
makes no contribution at all, and it suffices to use just the first
term of the above expansion, which is written explicitly.
We deduce that $\rho_0$ is
a singular point if and only if the following equations hold:
\[ \frac{\partial\det \rho}{\partial\rho_{kk}}(\rho_0)=
\lambda_1\lambda_2\cdots\hat{\lambda}_{k}\cdots
\lambda_{n-1}(\lambda_n-\lambda_k)=0,
\quad 1\le k\le n-1 \]
where the hat means that $\lambda_k$ should be omitted.
It is easy to see that these equations are satisfied if and only if
at least two of the eigenvalues are 0.
\end{proof}

Note that if $n=2$ then it is impossible for both eigenvalues to
be 0, which means that in this special case the hypersurface 
$\pD$ is smooth. Indeed, we have $\partial\pM=\pD$, $\pD$ is
a shere, and $\pM$ the corresponding solid 3-dimensional ball.

The following corollary follows immediately from the proposition.

\begin{corollary}
A point $\rho_0\in\pD^\Gamma$ is a singular point of $\pD^\Gamma$
if and only if $\rho_0^\Gamma$ has at least two zero eigenvalues.
\end{corollary}

Let us return now to the case of two qubits.
In that case we shall prove that the piece $\pD^\Gamma\cap\pM^0$,
where $\pM^0$ denotes the relative interior of $\pM$, contains no
singular points of the hypersurface $\pD^\Gamma$, i.e., for each
$\rho\in\pD^\Gamma\cap\pM^0$, the operator $\rho^\Gamma$
has exactly one zero eigenvalue.

\begin{theorem}
The piece of the hypersurface $\pD^\Gamma$ contained in the
(relative) interior of $\pM$ is smooth.
\end{theorem}

\begin{proof}
Assume that a point $\rho_0\in\pD^\Gamma\cap\pM^0$ is a singular
point of $\pD^\Gamma$. Then $\rho_0^\Gamma$ is a singular point
of $\pD$ and so it has at least two zero eigenvalues.
We shall now apply an argument of Sanpera et al. \cite{STV}.
By their Theorem 2, the zero eigenspace of $\rho_0^\Gamma$
contains a product vector $|e,f\ra$, and obviously we have
\[ \la e,f| \rho_0^\Gamma | e,f\ra =0. \]
But this expression is equivalent to
\[ \la e,f^* | \rho_0 | e,f^* \ra = 0, \]
which is impossible since $\rho_0\in\pM^0$, and so $\rho_0>0$.
This contradiction proves the theorem.
\end{proof}

The following corollary is obvious.

\begin{corollary}
The piece of the hypersurface $\pD$ contained in the
(relative) interior of $\pM^\Gamma$ is smooth.
\end{corollary}

\end{document}